\documentclass[12pt]{article}
\begin{document}
\begin{large}
\title{\bf {On probability, indeterminism and quantum paradoxes}}
\end{large}
\author{Bruno Galvan \footnote{Electronic address: b.galvan@virgilio.it}\\ \small Loc. Melta 40, 38014 Trento, Italy.}
\date{\small April 2004}
\maketitle
\begin{abstract}
If the quantum mechanical description of reality is not complete and a hidden variable theory is possible, what arises is the problem to explain where the rates of the outcomes of statistical experiments come from, as already noticed by Land\'e and Popper. In this paper this problem is investigated, and a new ``paradigm" about the nature of dynamical and statistical laws is proposed. This paradigm proposes some concepts which contrast with the usual intuitive view of evolution and of physical law, such as: initial conditions could play no privileged role in determining the evolution of the universe; the statistical distribution of the particles emitted by a source could depend on the future interactions of the particles; indeterministic trajectories can be defined by the least action principle.

This paradigm is applied to the analysis of the two-slit experiment and of the EPR paradox, and a coherent picture for these phenomena is proposed; this new picture shows how the well known difficulties in completing the quantum mechanical description of reality could be overcome.
\vspace{5 mm}
\end{abstract}
%\begin{large}
%newpage
\section{Introduction}

The possibility that the wave function (or quantum state) could not be the complete description of a physical system has been widely debated since the famous paper by Einstein, Podolsky and Rosen {\it Can Quantum Mechanical Description of Reality Be Considered Complete?} \cite{epr}. The search for a hidden variable description that could substitute the wave function description has been contrasted by some no-go theorems, the most effective of which is the EPR paradox \cite{bell}. One of the assumptions on which the EPR paradox is based is the so called {\it independence assumption} (IA), introduced and widely discussed by Price \cite{price2}\cite{price3}\cite{price4}; one of its many possible formulations is: 
\begin{quote} 
{\it The statistical distribution of the variables (direction, energy, hidden variables,...) of the particles emitted by a source does not depend on the future interactions of the particles.}
\end{quote}
This assumption seems so obvious that in the first proofs of the paradox it was not even mentioned, and only in more recent times has it been officially included in the hypotheses of the paradox \cite{valdenebro}. IA is also implicitly assumed (but not yet officially recognized) in the analysis of the two-slit experiment, as we will see in section 4.

Price criticizes IA on the base of its evident time asymmetry. He suggests that its possible violation is higtly counterintuitive for us due to the fact that IA surely holds for the macroscopic world of our direct experience, but nothing forbids its violation at microscopic level.

\vspace{3mm}
The original motivation of the present research was to investigate the nature and the validity of the IA. This investigation however has lead to study the origin of the rates of the outcomes in statistical experiments, of course in a universe without wave function but with variables, hidden or not. This study should explain for instance why, by tossing a coin, we obtain 50\% heads and 50\% tails; why in a scattering experiment incident particles should be uniformly distributed on the incidence plane; where, in the EPR paradox, the distribution $\rho(\lambda)$ of hidden variables  comes from, and so on.

There are very few studies on this issue in the literature. To my knowledge, the only authors who have discussed this problem are Land\'e \cite{lande} and Popper \cite{popper}. Their conclusion is that in a deterministic universe statistical laws cannot be derived only from the dynamics, but a further element is needed, something like a statistical distribution (a measure) on the initial conditions of the universe. This conclusion however leads these authors to reject determinism.

In this paper the need for a measure to explain statistical laws is accepted, clarified and generalized, and these concepts are collected in what I have called the {\it trajectory paradigm}, or t-paradigm. Two important features of this paradigm are that it allows (1) the violation of IA and (2) the definition of indeterministic trajectories. It is due to these features that t-paradigm could overcome the usual difficulties of hidden variable theories, and allow a description of reality not based on the wave function and its reduction.

\vspace{3mm}
The plan of the paper is the following. In section 2 the conceptual bases of the t-paradigm are given. In section 3 there are some examples of systems based on the t-paradigm. In section 4 the two-slit experiment is discussed, and the possibility is proposed that it could be the experimental evidence of the violation of IA at microscopic level. In section 5 an idealized model of the universe is proposed, in which particles travel along classical trajectories, but quantum interference phenomena could be present. In section 6 two examples are given of indeterministic processes represented by systems of indeterministic trajectories: the decaying particle and the spin 1/2 particle in the Stern-Gerlach apparatus. In section 7 the EPR paradox is discussed. In section 8 a possible contribution of the t-paradigm to cosmology is discussed. In section 9 the problem of free will in the t-paradigm is shortly discussed.

%newpage
\section{Conceptual bases of the trajectories paradigm}
To expose the t-paradigm I'll define first the following simple mathematical structure: a {\it trajectories system} (t-system) is the system $ (M,\Lambda,\cal F, \mu)$, where $M$ is a generic set said the configuration space, $\Lambda$ is a set of trajectories $ \lambda : R \rightarrow M $, $ \cal F $ is a $\sigma $-algebra on $ \Lambda $ and $ \mu $ is a measure on $ \cal F $. A t-system is said {\it deterministic} if whenever there exist two trajectories $\lambda_1$ and $\lambda_2$, two times $t_1$ and $t_2$, and $\epsilon>0$ so that $\lambda_1(t_1+t)=\lambda_2(t_2+t)$ for $0\leq t\leq\epsilon$, then $\lambda_1 (t_1+t)=\lambda_2(t_2+t)$ for all $t \in R$. A {\it semi-trajectories system} (st-system) is the system $ (M,\Lambda^+,\cal F, \mu)$, where everything is like a t-system with the exception of $\Lambda^+$, which in this case is a set of semi-trajectories $\lambda^+:R^+\rightarrow M$. The definition of determinism can be easily extended to an st-system. 

The proposal of the t-paradigm is that the universe (including its quantum phenomena) could be represented as a t-system or an st-system. Actually, a t-system can only represent a flat, non relativistic idealized universe; this is however enough to expose and explain the conceptual bases of the t-paradigm. Then it will be possible to apply these concepts to more realistic models of the universe. It is important to specify that the t-system must represent the {\it whole} universe, not a part or a subsystem of it.

\vspace{3mm}
For a universe composed by $N$ stable interacting particles, the configuration space is $R^{3N}$ as usual. In section 6, other kinds of configuration spaces for decaying particles and for particles with spin will be considered.

The set $\Lambda$ is the set of all the dynamically admissible trajectories of the universe, and it should explain the dynamical laws of nature, such as the motion of bodies and the energy-momentum conservation. Trajectories can be deterministic as well as indeterministic, and no request or constraint is made about how they are defined; in particular, no request is made that they are built applying a law of motion to a certain set of initial conditions. Actually, in all the proposed examples, deterministic as well as indeterministic trajectories will be defined by the least action principle plus, if necessary, some structural constraints.

\vspace{3mm}
As already noticed by Land\'e and Popper, in a deterministic universe statistical phenomena -- i.e. the rates of the outcomes in statistical experiments-- cannot be explained by a pure dynamical law. I propose to generalize this result to a universe which admits deterministic as well as indeterministic trajectories. The reasoning is the following; suppose that: (1) a really performed experiment always gives certain rates for its outcomes; (2) it is reasonable to accept that, among the dynamically admissible trajectories (deterministic or indeterministic), there exist some trajectories for which that experiment gives different rates, also if these rates are never observed (for instance, it is reasonable to accept that there exist a dynamically admissible trajectory in which the rate of heads in long series of tosses of fair coins is not $1/2$). Hence dynamical laws alone cannot predict the observed rates and another kind of law is needed; this law must state in some way that trajectories with the observed rates are ``much more" that the wrong ones. A precise example of this reasoning is the Bernoulli system described in the next section.

This new kind of law is a measure on $\Lambda$, and it works as follows. Consider a yes/no experiment; it will be described by a set of instructions which say how to prepare the experimental setting, how to perform the experiment and when the yes outcome occurs. Given this set of instructions, one can apply it to the trajectories of the set $\Lambda$ and, for every trajectory, get out the number $N$ of times the experiment has been performed in that trajectory and the rate $f$ of the yes outcome. Once fixed a large number $N$, let $\Lambda_{N}$ denote the set of trajectories for which the experiment is performed more than $N$ times, and assume that $\mu(\Lambda_{N})$ is finite, so that it can be normalized to $1$. Therefore an experiment defines a map $f:\Lambda_N\rightarrow [0,1]$, where $f(\lambda)$ is the rate of the yes outcome in the trajectory $\lambda$. Let $\langle f \rangle$ and $\Delta f$ denote the expectation value and the variance of the function $f$, defined in the usual way:
\begin{equation}
\langle f \rangle=\int_{\Lambda_{N}}{f(\lambda)d\mu},
\end{equation}
\begin{equation}
\Delta f=\langle f^2 \rangle-\langle f \rangle^2.
\end{equation}

If $\Delta f \ll 1$, then ``almost all" the trajectories have a rate near $\langle f\rangle$, and one can infer these two conclusions: (1) the experiment is well defined, i.e. its definition includes all the relevant conditions for the experiment and the experiment is repeatable; (2) $\langle f\rangle$ is the rate of the yes outcome of the experiment, that is its probability. This reasoning can be easily generalized to an experiment with $n$ possible outcomes.

\vspace{3mm}
A few considerations are now proposed. (a) The measure $\mu$ has extraordinary effects on the observed phenomena, because it completely determines the probabilities of the outcomes of the statistical experiments as well as their relevant conditions (note that given any surjettive map $f:\Lambda_N\rightarrow [0,1]$ and any expectation value and variance for $f$, there can be found a measure that gives rise to that expectation value and that variance). Therefore, by changing the measure one can obtain other rates for the outcomes of an experiment, or change a well defined experiment into a not well defined one (and vice versa). For instance, by suitably choosing the measure, one could obtain a universe in which the head outcomes of coin tosses have a rate of 60\% whenever the coins are tossed by a blue-eyed person; or, in experiments in which particles emitted by a source hit a screen, we could obtain on the screen any kind of figure, which may even depend on the weather! Therefore, one can understand that not only is the violation of IA possible, but it is just one of the innumerable counterintuitive effects which may be brought about by a suitable choice of the measure $\mu$.

(b) The deterministic or indeterministic character of the trajectories is irrelevant in this explanation of statistical laws.

(c) Here too, no {\it a priori} request is made about how the measure $\mu$ is defined. However, in the proposed examples, all the measures on the trajectories will be defined by means of measures on their boundary conditions (but not always initial conditions), and these measures will be either Lebesgue measures or measures derived by spectral measures on Hilbert spaces.

(d) The measure $\mu$ is defined on the trajectories as a whole, and it is not possible, in general, to attribute a definite measure to a set of trajectories of a subsystem of the universe. Therefore, according to the t-paradigm, the statistical independence between a measured system and the measuring apparatus, also before their interaction, is surely not a fundamental feature of the universe. This is another way to view the origin of the violation of IA.

\vspace{3 mm}
It is important to repeat that the set $\Lambda$ and the measure $\mu$ can be defined independently and in an atemporal way, i.e. considering the possibility of imposing global constraints and giving no privileged role to initial conditions with respect to other kinds of boundary conditions. The only obvious requests for $\Lambda$ and $\mu$ are that they should be defined in a simple way and, of course, that they should explain observed phenomena.

As already said, the true goal of the t-paradigm is to explain quantum phenomena; its way of defining statistical laws, its possibility of violating IA and of defining indeterministic trajectories are powerful tools that could overcame the well known problems of hidden variables theories. In this paper some steps in this direction will be taken.

%newpage
\section{Some examples of t-systems}

In all the examples of this paper, $\sigma$-algebras and measures will be defined on $\Lambda$ by transferring them from sets of boundary conditions for the trajectories. More precisely: suppose that there is a map $h_a$ from $\Lambda$ to a set $X_a$, not necessarily bijective; here $X_a$ is the boundary conditions set, and it is endowed with a $\sigma$-algebra ${\cal F}_a$ and a measure $\mu_a$. Then, $\sigma$-algebra and measure can be pulled back from $X_a$ to $\Lambda$ by defining ${\cal F}:=\{h^{-1}_a(\Delta_a)|\Delta_a\in {\cal F}_a\}$ and $\mu(\Delta):=\mu_a[h_a(\Delta)], \, \Delta\in {\cal F}$. Moreover, suppose that another boundary conditions set $X_b$ exists, with a $\sigma$-algebra ${\cal F}_b$ defined on it; suppose also that the map $h_b:\Lambda\rightarrow X_b$ is measurable. Then the measure $\mu$ on $\Lambda$ can be transferred on $X_b$ by defining $\mu_b(\Delta_b):=\mu[h_b^{-1}(\Delta_b)]$, $\Delta_b\in {\cal F}_b$. Therefore, the measure $\mu_a$ on $X_a$ can be transferred on $X_b$.

In discussing the examples, I will often say that a distribution on $X_a$ corresponds to another distribution on $X_b$ (or vice versa); by ``distribution" I mean measure, and by ``corresponds" I mean that the measure is transferred to the other space through the above defined method.
%newpage
\subsection{The Bernoulli system}
The configuration space M for this st-system is the semi-open interval $[0,1)$; in this example trajectories are maps from N (instead of from $R^+$) to $M$, and they are defined by repeated applications of the Bernoulli map $G(x)=2x$ (mod 1), i.e. $\lambda_x(n)=G^n(x)$, $x\in [0,1)$. Therefore $\Lambda=\{\lambda_x|x\in [0,1)\}$. The action of the Bernoulli map can be easily visualized: the binary representation of the numbers of the interval $[0,1)$ has $0$ as integer part and has a sequence of $0's$ and $1's$ as ``decimal" part. The action of the Bernoulli map is to shift the decimal part one place to the left; if the first element of the decimal part is $1$, the number is $\geq 1/2$. Let $X_a$ be the interval $[0,1)$, defined by the natural bijection $h_a:\lambda_x\mapsto x$. On $X_a$ the Borel $\sigma$-algebra and the Lebesgue measure are defined, and it can be considered as the set of initial conditions.

For this system the following yes/no experiment can be defined: the experiment takes place every step $n$, and its outcome is yes on a trajectory $\lambda$ at the step $n$ if $\lambda(n)\geq 1/2$. The rate $f$ of the yes outcome is defined by the limit:
\begin{equation}
  f(\lambda)=\lim_{n\rightarrow \infty}\frac{C(\lambda,n)}{n},
\end{equation}
where $C(\lambda,n)$ is the number of yes outcomes in the first $n$ steps of the trajectory $\lambda$, with the convention that $f(\lambda)=0$ if the limit (3) does not exist. It can be shown that the subset of trajectories for which the rate $f$ differs from $1/2$ has null measure (see for instance \cite{probability}). As a consequence, $\langle f\rangle=1/2$ and $\Delta f=0$, the experiment is well defined and the probability of the yes outcome is $1/2$. It can also be shown that, once fixed any value between $0$ and $1$, a measure can be found on $X_a$, which measure gives rise to that value for $\langle f\rangle$.

As to this system, it is very easy to show that the rate $1/2$ cannot be deduced only from the dynamical set $\Lambda$. Indeed, there exist trajectories for which $f(\lambda)\ne 1/2$; consider for instance the initial condition $2/7$, which gives rise to the rate $f=2/3$.

%newpage
\subsection{The scattered particle}
Consider a particle subjected to a repulsive potential $V(r)$, with $V(r)\rightarrow 0$ as $r\rightarrow \infty$. Let $M=R^3$ be the configuration space. Moreover, let $\Lambda$ be the set of trajectories satisfying the equation ${\bf f}=m{\bf a}$ (or equivalently, the least action principle) and having fixed energy $E>0$ and fixed incidence direction $\hat{u}_I\in S^2$. Every trajectory of $\Lambda$ defines the following two kinds of boundary conditions: $X_a:=\{{\bf s}|{\bf s}\in R^2\}$, where ${\bf s}$ is the (vector) impact parameter, and $X_b:=\{\hat{u }_S |\hat{u}_S \in S^2\}$, where $\hat{u}_S $ is the scattering direction. The sets $X_a$ and $X_b$ are endowed with Borel $\sigma$-algebras and Lebesgue measures as usual. A bijection exists between $X_a$ and $X_b$, therefore measures can be transferred from $X_a$ to $X_b$ and vice versa. Let us consider the coordinates $(s,\phi)$ on $X_a$ and $(\theta,\phi)$ on $X_b$, where $s\geq 0$ is the impact parameter and $-\pi < \theta\leq \pi$ is the scattering angle. If $d\mu_a(s,\phi)=\rho_a(s,\phi)\, s\, ds\, d\phi $ and $d\mu_b(\theta,\phi)= \rho_b(\theta,\phi)\sin\theta \, d\theta \, d\phi $ are two corresponding measures on $X_a$ and $X_b$, then the following equation relates the densities $\rho_a(s,\phi)$ and $\rho_b(\theta, \phi)$:
\begin{equation}
\rho_b(\theta,\phi) =\rho_a(s,\phi) \frac{s}{ \sin\theta }\left|\frac{\partial s}{\partial\theta}\right|.
\end{equation}

Conditions $X_a$ can be considered as the initial conditions. The deriving of the measure $\mu$ from the Lebesque measure on $X_a$ corresponds to the usual assumption that incident particles are uniformly distributed on the incident plane; this assumption is used in the derivation of the classical scattering cross-section. On the contrary, the deriving of $\mu$ from the Lebesgue measure on $X_b$ corresponds to assume that particles are uniformly scattered in all the directions. This uniform distribution of the scattering direction corresponds to no uniform distribution of ${\bf s}$, which also depends on the potential; this is an example of violation of IA. In a universe with an $X_b$-like measure, in scattering experiments the distribution of incident particles would depend on the target, and the particles would be uniformly diffused in all the directions! Surely our universe has not an $X_b$-like measure, but are we sure that it has an $X_a$-like measure? In section 5 I'll propose a third possibility.

%newpage
\subsection{The scattering cross-section}
The previous example is too simple to define an experiment and to assign rates of outcomes to the trajectories. Let us consider the following example: a large number of identical repulsive interaction centers $V(|x-x_k|)$ with action range $r_0$ is uniformly distributed in a large region $R$, and the distance between two centers is much greater than $r_0$. Trajectories intersecting the region $R$ are scattered many times by the interaction centers (like balls in a flipper).

An experiment with $n$ possible outcomes can be defined as follows: an experiment takes place on a trajectory if the trajectory passes at a distance less than $r_0$ from an interaction center, and the $i$-th outcome is obtained if the scattering angle of the interaction is in between $[-\pi +2\pi(i-1)/n]$ and $[-\pi +2\pi i/n]$, where $i=1,...,n$. Only trajectories with a number of experiments greater than a suitable number $N$ will be considered. Every such trajectory defines the set of rates of the outcomes $\{f_i(\lambda)\}$; this set can be considered as the {\it scattering cross-section} defined by the trajectory (actually the right cross-section is $\{f_i \pi r_0^2\}$). A definite cross-section will exist for this universe if $\Delta f_i \ll 1$ for all $i$; in this case the cross-section will be $\{\langle f_i\rangle\}$.

I think that also without proof, it is easy to accept that: (i) there exist trajectories with a cross section close to any given cross-section; (ii) given any cross-section $\{\bar{f_i}\}$, there can be found a measure $\mu$ so that $\{\langle f_i\rangle\}=\{\bar{f_i}\}$; (iii) the type $X_a$ measure of the previous example, corresponding to the uniform distribution of the incident particles, gives rise to the classical cross-section.

%newpage
\subsection{The classical big bang model}
This example of st-system represents an idealized newtonian universe with a ``big bang".

Consider a system of $N$ particles interacting through everywhere bounded central potentials. Let $\Lambda^+$ be the set of classical semi-trajectories for which $\lambda(0)=0$, that is for which, at the time $t=0$, all the particles are concentrated in the point $x=0$. The asymptotic behavior of the trajectories is governed by a theorem of classical scattering theory, which states that, with suitable (and very acceptable) conditions for the potentials, the limit:
\begin{equation}
\lim_{t\rightarrow+\infty}\frac{\lambda(t)}{t}=v^+_{\lambda}
\end{equation}
does exist for every trajectory $\lambda$, and its value is said {\it asymptotic velocity} \cite{scattering}. Qualitatively, it happens that particles collect themselves into bounded clusters which move far away from each-other, and whose velocities tend to a limit \cite{scattering2}.

Let $X_a=R^{3N}$ be the set of momenta of the trajectories at time $t=0$; the map $h_a:\Lambda^+\rightarrow X_a$ is bijective and, as usual, Borel $\sigma$-algebra and Lebesgue measure on $X_a$ can be transferred to $\Lambda^+$.

These boundary conditions can be considered as the initial conditions of the universe, and the induced measure corresponds to the uniform distribution of initial momenta. It is reasonable to assume that IA does hold in this universe.

The following boundary conditions can also be defined; suppose that a final time $T$ exists for the universe, and define $X_b(T)=\{\lambda(T)\}_{\lambda\in \Lambda^+}\subseteq R^{3N}$; note that the map $h_b:\Lambda^+\rightarrow\ X_b(T)$ is in general not bijective.

These are the boundary conditions of the least action principle. If initial momenta have a uniform distribution (i.e. a Lebesgue measure), boundary conditions $X_b(T)$ will have a very complicated and twisted distribution. One could also assume uniform distribution for boundary conditions $X_b(T)$; in this case boundary conditions $X_a$ will have a very complicated and twisted distribution; as a consequence, very strange phenomena would happen in this universe, such as, for instance, the violation of IA at macroscopic level.

Also asymptotic velocities $\{ v^+_{\lambda}\}_{\lambda\in \Lambda^+}$ are boundary conditions, and let $X_c$ denote this set. Of course, also the map $h_c:\Lambda\rightarrow X_c$ is not bijective. Boundary conditions $X_c$ do not need a final time for the universe, and they can be considered somehow as the limit for $T\rightarrow\infty$ of boundary conditions $X_b(T)$.

%newpage
\section{The two-slit experiment }

The main proposal of the present work is that the motion of the particles could be represented by trajectories (which for instance satisfy the least action principle) instead of by wave functions. It is then necessary to review the reasons that compelled the founders of quantum mechanics to use wave function not only to calculate the probability of finding the particles in a certain spatial region, but also to describe their motion. In other words, they claimed that a particle emitted by a source at $(t_1,\bf{x}_1)$ and revealed by a counter at $(t_2,\bf{x}_2)$ does not travel along the least action path which joins the two points, but it dematerializes in a wave after its emission and it materializes again in a particle when it is revealed by the counter; this is the {\it wave-particle dualism}, one of the most intriguing puzzles of quantum mechanics. The main reason for all this is the two slit experiment and its interference phenomena, as explained for instance by Heisenberg \cite{heisenberg}.

Let us consider the version of the experiment performed with electrons:
\begin{center}
\unitlength=1mm
\begin{picture}(120,35)
\put(54,5){\line(1,0){10}}
\put(54,30){\line(1,0){10}}
\put(58,17){\circle*{1}}
\put(112,5){\line(0,1){23}}
% sorgente
\put(0,16){\line(1,0){7}}
\put(7,16){\line(0,1){3}}
%\put(7,18){\line(0,1){1}}
\put(0,19){\line(1,0){7}}
\put(0,16){\line(0,1){3}}
\bezier{600}(7,18)(58,22)(112,17)
\bezier{600}(7,17)(58,13)(112,17)

\put(47,2){\makebox(6,6){$E_2$}}
\put(47,27){\makebox(6,6){$E_1$}}
\put(50,14){\makebox(6,6){$F$}}
\put(105,1){\makebox(6,6){$H$}}

\put(0,10){\makebox(6,6){$S$}}

\end{picture}

Fig. 1
\end{center}

Here $S$ is an electron source, $F$ is a tiny conducting wire which crosses the plane of the figure at right angle and can be set to a positive potential with respect to the two electrodes $E_1$ and $E_2$; $H$ is a screen constituted by a photographic plate. Due to the electrostatic field generated by the wire, the electrons emitted by the source are deflected and produce interference fringes on the screen. If the electrostatic field is turned off, the interference fringes disappear.

In a description of the motion with classical trajectories, the point of the screen hit by an electron depends on the little angle with which the electron is emitted by the source, and the figure on the screen depends on the statistical distribution of this angle. Maintaining the trajectory description, one must accept the fact that the statistical distribution of this angle changes when the electrostatic field is turned on and off, even if no interaction exists between the source and the deflecting device. Moreover, the variation is so microscopically precise that it gives rise to the correct interference fringes on the screen. This possibility seemed to physicists to hard to accept, and they preferred to abandon the trajectory representation and to use the wave function also to describe the motion of the particles. But what physicists have rejected could be just the experimental evidence of the violation of IA, which is absolutely allowed in the t-paradigm.

What kind of measure on the trajectories of the universe could determine the strange behavior of the distribution of the electrons emitted by the source? In the next section an unavoidably schematic model of universe with such a measure is proposed.

%newpage
\section{The quantum big bang model}

This model of universe has the same dynamical structure $\Lambda^+$ of its classical counterpart described in section 3. To define the quantum-like measure, we use the $X_b(T)$ boundary conditions; if $\Delta$ is a measurable subset of $X_b(T)$, we define: 
\begin{equation}
\mu_Q(\Delta)=\langle 0|U^+(T)E(\Delta)U(T)|0\rangle,
\end{equation}
where $|0\rangle$ is the eigenstate of eigenvalue $0$ of the $3N$ quantum position operators $Q=(\bf {Q}_1,...,\bf {Q}_N)$, $U(\cdot)$ is the quantum time-evolution operator and $E(\cdot)$ is the spectral measure of $Q$. In other words, the measure of $\Delta$ is the quantum probability to find the particles of the universe in the region $\Delta$ at the time $T$, being the universe in the state $|0\rangle$ at the time $t=0$.

In quasi-classical approximation the measure $\mu_Q$ can be explicitly calculated, and the result is:
\begin{equation}
\rho_Q(x)=\rho_C(x)+I(x), \: x\in X_b
\end{equation}
where $\rho_Q$ is the density of the measure $\mu_Q$, $\rho_C$ is the density of the measure on $ X_b(T)$ which corresponds to the uniform distribution of initial momenta (the $X_a$ measure of the classical big bang model), and I(x) is an interference term not always positive. The detailed calculation can be found in \cite{semiclassical}. It is reasonable to claim that the classical part $\rho_C$ gives rise to the macroscopic predominance of IA, and that the quantum interference term determines interference phenomena with violation of IA at microscopic level.

The quantum measure can also be defined on the boundary conditions $X_c$, i.e. on the asymptotic velocities. Similarly to the classical case, a theorem states that the limit 
\begin{equation}
\lim_{t\rightarrow+\infty}\frac{U^+(t)QU(t)}{t}=V^+.
\end{equation}
does exist, and that the $V^+$ are $3N$ operators which commute between each other and with the hamiltonian \cite{scattering}. Hence, if $\Delta$ is a measurable subset of $X_c$, it is natural to define
\begin{equation}
\mu_Q(\Delta)=\langle 0|F(\Delta)|0\rangle,
\end{equation}
where $F(\cdot)$ is the spectral measure of the operators $V^+$.

In absence of interactions, it can be shown that $V^+=P/m$, where $P$ are the momentum operators and $P/m=({\bf P}_1/m_1,\cdots, {\bf P}_N/m_N)$; the measure $\mu_Q$ becomes
\begin{equation}
\mu_Q(\Delta)=\frac{\mu_L(m\Delta)}{2\pi^{3N}},
\end{equation}
where $m\Delta:=\{(m_1{\bf v_1},\cdots, m_1{\bf v_1})|({\bf v_1},\cdots, {\bf v_1})\in \Delta\}$ and $\mu_L$ is the Lebesgue measure. Therefore, in absence of interactions, measure (9) corresponds to the uniform distribution of initial momenta.

%newpage
\section{Indeterministic trajectories}

In this section, by using the least action principle, two examples of systems of indeterministic trajectories are defined. The least action principle and the equations of motion are usually considered as equivalent tools to define trajectories, and it is interesting to notice that for indeterministic trajectories only the least action principle does work.

%newpage
\subsection{Decaying particle}
Consider a system composed by one free particle of mass $m_1$ which decays in two free particles of mass $m_2$ and $m_3$, with $m_1> m_2+m_3$. The configuration space of this system is $M_1\cup (M_2 \times M_3)$, where $M_i=R^3$ is the usual configuration space of a particle. The set $\Lambda$ of dynamically allowed trajectories is defined by the following three requirements: (1) for every trajectory $\lambda$ there exists a decaying time $t_d$ so that $\lambda(t)\in M_1$ for $t\leq t_d$ and $\lambda(t)\in M_2\times M_3$ for $t>t_d$. (2) Particles 2 and 3 originate in the same spatial point in which particle 1 decays, i.e. $\lambda(t_d^+)=(\lambda(t_d),\lambda(t_d))$. (3) Between any two times $t_I<t_F$ the trajectory minimizes the action $S=\int_{t_I}^{t_F}{L(x,\dot{x})dt}$, where
\begin{equation}
L(\dot{x}, x)=\left\langle
\begin{array}{ll}
\frac{1}{2} m_1{\bf \dot {x}_1}^2 -m_1c^2 & for \: x={\bf x}_1 \in M_1 \\ 
\\
\frac{1}{2} m_2{\bf \dot {x}_2}^2 -m_2c^2 +
\frac{1}{2} m_3{\bf \dot {x}_3}^2 -m_3c^2 
& for \: x=({\bf x}_2, {\bf x}_3)  \in M_2\times M_3
\end{array}
\right.
\end{equation}
The lagrangian (11) is the approximation for low velocities of the free relativistic lagrangian $L=-mc^2\sqrt{1-\beta^2}$. When both $ \lambda(t_I)$ and $ \lambda(t_F)$ belong either to $M_1$ or to $M_2 \times M_3$, the trajectory is a straight line between the points $ \lambda(t_I)$ and $ \lambda(t_F)$. When $ \lambda(t_I)={\bf x}_1\in M_1$ and $ \lambda(t_F)=({\bf x}_2, {\bf x}_3) \in M_2\times M_3$, the action can be written as follows:
\begin{eqnarray}
S & = & -m_1c^2(t_d-t_I)+ \frac{1}{2}m_1\frac{({\bf x}_d-{\bf x}_1)^2}{t_d-t_I}
-m_2c^2(t_F-t_d)+ \frac{1}{2}m_2\frac{({\bf x}_2-{\bf x}_d)^2}{t_F-t_d} \\
& & -m_3c^2(t_F-t_d)+ \frac{1}{2}m_3\frac{({\bf x}_3-{\bf x}_d)^2}{t_F-t_d}.\nonumber
\end{eqnarray}
where $(t_d,{\bf x}_d)$ is the time/point in which particle 1 decays.
The action is minimized by equating its derivatives with respect to $(t_d,{\bf x}_d)$ to 0. This gives:
\begin{equation}
m_1\frac{{\bf x}_d-{\bf x}_1}{t_d-t_I}
-m_2\frac{{\bf x}_2-{\bf x}_d}{t _F-t_d}
-m_3\frac{{\bf x}_3-{\bf x}_d}{t_F-t_d }=0,
\end{equation}
\begin{equation}
-m_1c^2-\frac{1}{2}m_1\frac{({\bf x}_d-{\bf x}_1)^2}{(t_d-t_I)^2}+
m_2c^2+\frac{1}{2}m_2\frac{({\bf x}_2-{\bf x}_d)^2}{(t_F-t_d)^2}+
m_3c^2+\frac{1}{2}m_3\frac{({\bf x}_3-{\bf x}_d)^2}{(t_F-t_d)^2}=0,
\end{equation}
which are the equations of momentum and energy conservation. Equations (13) and (14) form a system of four equations which allow to compute the four unknown variables $({\bf x}_d,t_d)$ (the detailed calculation is not given here). It is obvious that the trajectories so defined are indeterministic, and no equation of motion is able to fix the future evolution of particle 1 given its position and its momentum at time $t_I$.

Of course the only definition of the set $\Lambda$ does not say anything about the mean life of the decaying particle, which is a statistical law and needs a measure to be defined. Let us consider then the set composed by the positive-time part of the trajectories of the set $\Lambda$ for which $\lambda(0)=0\in M_1$; let $\Lambda^+$ denote this set. Let the set $\Lambda$ be also endowed with a measure $\mu$ so that $\mu(\Lambda^+)=1$. Than the mean life of particle 1 is simply
\begin{equation}
\int_{\Lambda^+}{t_d(\lambda)d\mu(\lambda)}.
\end{equation}

%newpage
\subsection{Spin 1/2 particle in a magnetic field}
Let $M\times C^2=\{({\bf x},s)| {\bf x}\in R^3, s\in C^2\}$ be the configuration space of a spin 1/2 particle, where $s$ is the spin variable, which must not be confused with the quantum spin state, even if they have the same mathematical representation. The hamiltonian for a neutral quantum spin 1/2 particle in a magnetic field (for instance an hydrogen atom) is the Pauli equation:
\begin{equation}
H=\frac{P^2}{2m} + \mu {\bf \hat{\sigma}}\cdot {\bf B},
\end{equation}
where
\begin{equation}
\mu=2\frac{e}{2m_e c}\frac{ \hbar}{2}
\end{equation}
and $\hat{\sigma}=(\sigma_1, \sigma_2, \sigma_3)$ are the Pauli matrixes. From analogy, I propose that the trajectories of a spin 1/2 particle in presence of a magnetic field must minimize the action given by the lagrangian
\begin{equation}
L=\frac{1}{2}m\dot{\bf x}^2 -\mu\frac{\langle s|\hat{\sigma}\cdot {\bf B}|s\rangle}{\langle s|s\rangle}.
\end{equation}
The Eulero-Lagrange equations applied to the lagrangian (18) give:
\begin{equation}
{\bf \sigma}\cdot {\bf B}|s\rangle=\frac{\langle s|\hat{\sigma}\cdot {\bf B}|s\rangle}{\langle s|s\rangle}|s\rangle,
\end{equation}
\begin{equation}
m\ddot{\bf x}=-\mu\nabla_{\bf x}\left(\frac{\langle s| \hat{\sigma}\cdot {\bf B}|s\rangle}{\langle s|s\rangle}\right).
\end{equation}
Note that equation (19) is not a true equation of motion, because of the absence of a time derivative term for $s$; this will be the cause for the indeterministic character of the trajectories. From equation (19) we obtain that the spin variable $s$ must be an eigenstate of $\hat{\sigma}\cdot {\bf B}$, whose eigenvalue are $\pm |{\bf B}|$; it follows that (19) and (20) become:
\begin{equation}
{\bf \sigma}\cdot {\bf B}|s\rangle =\pm |{\bf B}||s\rangle
\end{equation}
\begin{equation}
m\ddot{\bf x}=\mp \mu\nabla_{\bf x}|{\bf B}|.
\end{equation}
Consider now the Stern-Gerlach experiment.
\begin{center}
\unitlength=1mm
\begin{picture}(130,35)

\put(40,20){\line(1,0){50}}
\put(40,20){\line(0,1){10}}
\put(90,20){\line(0,1){10}}

\put(40,10){\line(1,0){50}}
\put(40,0){\line(0,1){10}}
\put(90,0){\line(0,1){10}}

\put(10,15){\circle*{1}}
\put(120,5){\circle*{1}}
\put(120,25){\circle*{1}}

\put(7,17){\makebox(6,6){$S$}}
\put(123,2){\makebox(6,6){$R_-$}}
\put(123,22){\makebox(6,6){$R_+$}}

\bezier{600}(25,15)(80,15)(120,25)
\bezier{600}(25,15)(80,15)(120,5)
\put(10,15){\line(1,0){15}}

\put(32,23){\makebox(6,6){$A$}}
\bezier{20}(35,7)(35,15)(35,22)

\put(92,23){\makebox(6,6){$B$}}
\bezier{20}(95,7)(95,15)(95,22)

\end{picture}

Fig. 2
\end{center}
A particle is emitted by the source $S$. In the part $S-A$ there is no magnetic field and, due to equation (22), the particle travels along a straight line; equation (21) imposes no constraint to the spin variable, so trajectories with any value for $s$ in $S-A$ are dynamically admissible. At point $A$ the particle meets the magnetic field, and instantaneously the spin variable orients itself along the direction of the field, but with impredictable sign. In the part $A-B$ the particle is influenced by the magnetic gradient according to the sign of the spin; therefore both paths $S-R_+$ and $S-R_-$ are allowed. This shows that the system is indeterministic, because the spin and the other variables of the particle near $S$ can be the same for both paths. Note that here too, dynamics says nothing about the probability with which either path will be chosen by the particle. More about this in the next section.

%newpage
\section{The EPR paradox}
In spite of the fact that the t-paradigm allows the violation of IA, the picture given by the t-paradigm of the EPR experiment is not based on this violation. On the contrary, what is violated is Einstein's reality criterium \cite{epr}.

As a matter of fact, let us consider again the Stern-Gerlach experiment, and let us suppose that the magnetic field and its gradient are along the positive direction of the $z$-axis, and that near the source $S$ the particle is in the quantum spin state $|z,+\rangle$. Then the particle will certainly travel along the path $S-R_+$. Notwithstanding this, in absence of magnetic field there is no correlation between the quantum spin state and the spin variable, which can evolve freely and assume any value. As a consequence, near $S$ the two trajectories $S-R_+$ and $S-R_-$ can have the same values for all their variables, and no element of physical reality corresponds to the certainty of the $R_+$ result.

Note however that, according to the t-paradigm, it is not right to affirm that near the point $S$ the particle is in the quantum spin state $|z,+\rangle$, and the attribution of a quantum state to the particle is just a representation of the effects of the global measure $\mu$ on the particle subsystem. In order to better explain this point, let us go a little bit more into the details of how the t-paradigm represents the EPR experiment.

Let us consider the usual experimental setting: a source $S$ emits pairs of spin 1/2 particles in singlet state towards two distant Stern-Gerlach measuring devices $A$ and $B$, with possible orientations $(a,a')$ and $(b,b')$ respectively. The orientations of the devices are independently chosen by random mechanisms or by the experimenters after the emission of the particles.

Let $\Lambda_0$ denote the subset of trajectories of the universe which corresponds to the emission and to the measurement of a single specific pair of particles in a specific experiment performed in a specific place at a specific time. $\Lambda_0$ can be divided into sixteen subsets $\Lambda_0(O_A,R_A,O_B,R_B)$, which corresponds to every possible combination of orientations and results; here $O_A=a,a'$ is the orientation of the apparatus $A$, $R_A=+,-$ is the result of the measurement in $A$, and the same holds for $B$. Let us define also $\Lambda_0(O_A,O_B):=\bigcup_{R_A,R_B}\Lambda_0(O_A,R_A,O_B,R_B)$. It is natural to define the sixteen conditional probabilities for the outcomes $R_A$ and $R_B$ as
\begin{equation}
p(R_A,R_B|O_A,O_B):=\frac{\mu[\Lambda_0(O_A,R_A,O_B,R_B)]}{\mu[\Lambda_0(O_A,O_B)]}.
\end{equation}
The global nature of the measure $\mu$ can give any value for these conditional probabilities and, in particular, it can give the quantum mechanical values. This quantum measure should be similar to the one described in the quantum big bang model. If equation (23) provides the quantum mechanical probabilities for every measurement, then, following the law of large numbers, ``almost all" the trajectories of $\Lambda$ will show rates of outcomes which correspond to those probabilities.

Note that near the source $S$ there is no magnetic field, and, as already explained, any value for the spin variables of the particles is allowed.

%newpage
\section{T-paradigm and cosmology}

Now a possible contribution of the t-paradigm to cosmology is discussed. Let us consider for instance the flatness problem: in order to explain the actual value of $\Omega$, 1 second after the big bang, the total mass density of the universe must have been equal to the critical mass density $3H^2/8\pi G$ (which gives rise to a geometrically flat universe) up to the 15-th decimal place. Of course this coincidence is considered unacceptable, and several mechanisms, such as inflaction, have been proposed to solve the problem.

This is only an example of the usual approach according to which the structure and the possible evolutions of the universe depend on its initial conditions. As we have seen, t-paradigm has a totally different approach. According to this approach, the structure of the universe is built in an atemporal way, i.e. global structural constraints with the same role as dynamical laws may exist, and initial conditions may not play a privileged role in defining the evolution of the universe. As to the flatness problem, let us suppose that we discover that the universe is really flat; then flatness could be posed as one of the laws which define the dynamically admissible trajectories of the universe, and this law should be considered on the same level of the least action principle. Than initial conditions would depend on the flatness law, and not vice versa. The assumption that flatness must derive from initial conditions could correspond, in the quantum big bang model, to the assumption that the quantum measure (9) on the asymptotic velocities must derive from a measure defined on initial momenta of the universe; this would be a real puzzle!

%newpage
\section{Time flow and free will}
If the universe is built in an atemporal way, i.e. without any privileged role for initial conditions, where do our feelings of the flowing of time and of free will come from? Another querie: who chooses the trajectory of the universe? These two queries are connected, and I'll propose two possible answers.

(1) God (or nature) has built the universe with all its trajectories and its measure; then, considering the measure, He has chosen a trajectory (as a whole, not only its initial conditions) and has said us to follow this trajectory. Then we must agree with Parmenide and Einstein that our feelings of the flowing of time and of free will are pure illusions (see Popper \cite{popper}). Note that in this case free will is absent even if indeterministic trajectories are allowed.

(2) God has built the universe with all its trajectories and its measure, but He has not chosen any trajectory. It is the universe itself, and we in it, that choose its trajectory at every measurement. Consider for instance a particle emitted by a source which can hit one of a set of counters. Before any counter clicks, no conscious being in the universe knows what the trajectory of the particle is; so we can think that the trajectory has not yet been chosen, and the choice is made when a counter clicks and a conscious observer becomes aware of this. The same reasoning holds for indeterministic trajectories: the choice between two indeterministic trajectories is made when the two trajectories become different in such a way that a conscious observer can distinguish them. One can think of the set $\Lambda$ as reducing itself at every measurement, while excluding the trajectories that are no more compatible with the state of knowledge of the conscious observers of the universe. Also the brain processes could be considered as measurements, and, as a consequence of every decision we make, a subset of the dynamically admissible trajectories is chosen while the others are excluded. This point of view can be considered somehow close to Wigner's point of view, according to which the reduction of the wave function is due to the intervention of a conscious observer \cite{wigner}; the difference is that in this case what is reduced is not the wave function but the set of admissible trajectories.
It is obious that this second view is totally compatible with our feelings of the flowing of time and of free will.

\vspace{3 mm}
The difference between these two answers rests however only at philosophical level, and no difference exists between them at physical or experimental level.

%newpage
\section{Conclusion}
The study of the origin of statistical laws in a non wave function based universe has lead to the definition of what I have called the {\it trajectories paradigm}, or t-paradigm. The main proposals of the t-paradigm are: (i) Two kinds of laws are needed to explain observed phenomena of the universe: dynamical laws, which are represented by the set $\Lambda$ of dynamically admissible trajectories, and statistical laws, which are represented by a measure on $\Lambda$. (ii) These two kinds of laws should be defined in an indipendent and atemporal way; for instance, no privileged role should be given to initial conditions with respect to other kinds of boundary conditions. (iii) IA may not be a fundamental feature of reality, and, at microscopic level, its validity or violation should only be matter of experimental check. (iv) If the trajectories representation of the particles is maintained, the interference phenomena of the two-slit experiment could be the experimental evidence of the violation of IA. (v) Deterministic as well as indeterministic trajectories are allowed in the t-paradigm, and their being so has nothing to do with statistical laws (vi) The least action principle (but not the equations of motion) allows the definition of indeterministic trajectories; examples have been given in the case of a decaying particle and of a spin $1/2$ particle in a magnetic field. (vii) This last example in particular shows that Einstein's reality criterium --on which the EPR paradox is based-- is not valid in the t-paradigm. (iix) In cosmology, t-paradigm proposes to change perspective as to the role the initial conditions of the universe play in determining its structure and evolution; as an example, the flatness problem has been discussed according to this new perspective. (ix) The t-paradigm should be compatible with free will.

Following the above mentioned possibilities, the t-paradigm could overcome the well known difficulties in completing the quantum mechanical description of reality, and allow a description of reality not based on the wave function.

%newpage

%\end{large}
\end{document}